\documentclass[letterpaper]{article} % DO NOT CHANGE THIS
\usepackage{aaai2026}  % DO NOT CHANGE THIS

\nocopyright
\usepackage{times}  % DO NOT CHANGE THIS
\usepackage{helvet}  % DO NOT CHANGE THIS
\usepackage{courier}  % DO NOT CHANGE THIS
\usepackage[hyphens]{url}  % DO NOT CHANGE THIS
\usepackage{graphicx} % DO NOT CHANGE THIS
\usepackage{natbib}  % DO NOT CHANGE THIS AND DO NOT ADD ANY OPTIONS TO IT
\usepackage{caption} % DO NOT CHANGE THIS AND DO NOT ADD ANY OPTIONS TO IT
\usepackage{algorithm}
\usepackage{algorithmic}
\usepackage{booktabs}
\usepackage{amsfonts}%
\usepackage{amsmath}
\usepackage{multirow}
\usepackage{tabularx}
\usepackage{adjustbox}
\newcolumntype{d}[1]{D{.}{.}{#1}}

\usepackage{newfloat}
\usepackage{listings}
\DeclareCaptionStyle{ruled}{labelfont=normalfont,labelsep=colon,strut=off} % DO NOT CHANGE THIS
\floatstyle{ruled}
\newfloat{listing}{tb}{lst}{}
\floatname{listing}{Listing}

\title{Generalizable speech deepfake detection via meta-learned LoRA}
\author{
    Janne Laakkonen\textsuperscript{\rm 1}\equalcontrib, 
    Ivan Kukanov\textsuperscript{\rm 2}\equalcontrib, 
    Ville Hautamäki\textsuperscript{\rm 1}
}
\affiliations{
    \textsuperscript{\rm 1}School of Computing\\
    University of Eastern Finland\\
    Joensuu, Finland\\
    \textsuperscript{\rm 2}KLASS Engineering and Solutions \\
    Singapore

}

\usepackage{bibentry}
\begin{document}

\maketitle

\begin{abstract}
Reliable detection of speech deepfakes (spoofs) must remain effective when the distribution of spoofing attacks shifts. 
We frame the task as domain generalization and show that inserting Low‑Rank Adaptation (LoRA) adapters into every attention head of a self-supervised (SSL) backbone, then training only those adapters with Meta‑Learning Domain Generalization (MLDG), yields strong zero‑shot performance. 
The resulting model updates about 3.6\,million parameters, roughly 1.1\,\% of the 318\,million updated in full fine‑tuning, yet surpasses a fully fine‑tuned counterpart on five of six evaluation corpora. 
A first‑order MLDG loop encourages the adapters to focus on cues that persist across attack types, lowering the average EER from 8.84\,\% for the fully fine‑tuned model to 5.30\,\% with our best MLDG–LoRA configuration. 
Our findings show that combining meta-learning with parameter-efficient adaptation offers an effective method for zero-shot, distribution-shift-aware speech deepfake detection.
\end{abstract}

\section{Introduction}
\label{sec:intro}

Machine‑learning models often struggle when they encounter data distributions that differ from those seen during training.  
\emph{Domain generalization} (DG) directly addresses this challenge: the goal is to train a model on source domains that performs effectively on \emph{unseen} target domains, provided the underlying task semantics remain consistent \cite{Zhou_2022}.  
Consider an image classifier trained to recognize elephants in photographs; true generalization means it can still identify a hand‑drawn elephant—despite vastly different pixel statistics—by focusing on invariant features such as the trunk and large ears rather than superficial stylistic elements.

Speech deepfake detection faces an analogous challenge.  
While obtaining large amounts of \emph{bonafide} speech is relatively simple, new \emph{spoofs} are constantly being created.  
Attackers can deploy entirely new text‑to‑speech (TTS) or voice‑conversion (VC) algorithms, or simply retrain an existing generator with a different random seed, instantly creating a novel acoustic distribution \cite{Wang2024ASVspoof5C}.  
Consequently, a detector trained only on known attack types is prone to failure when faced with these unseen synthesis methods.  
Reliable deepfake detection therefore fundamentally requires a DG approach: the model must learn acoustic cues of synthetic speech that generalize across attack types instead of merely memorizing the characteristics of those seen during training.

A common strategy for handling multiple attack types is \emph{Empirical Risk Minimization} (ERM) \cite{VapnikNIPS1991}, in which data from all known attacks are pooled into a single training set.  
Although ERM is often a strong baseline \cite{gouk2024limitationsgeneralpurposedomain}, it does not explicitly exploit the domain structure—that is, the distinct attack types—to optimize for generalization.  
Another direction fine‑tunes large, pre‑trained foundation models such as \texttt{Wav2Vec 2.0} \cite{Tak2022Automatic, wang2023lowrankadaptationmethodwav2vec2based}.  
These models capture rich speech representations, but fully fine‑tuning them can be computationally demanding, requiring considerable time and memory, and risks overfitting to the specific characteristics of the training attacks, potentially hindering performance on unseen ones \cite{müller2024doesaudiodeepfakedetection}.

To address these limitations, we investigate the synergy between meta‑learning for domain generalization and parameter‑efficient fine‑tuning. Our approach pairs Meta‑Learning Domain Generalization (MLDG) \cite{Li2017LearningTG}, which explicitly simulates domain shifts during training to improve generalization, with Low‑Rank Adaptation (LoRA) \cite{hu2021loralowrankadaptationlarge}, a technique for efficiently adapting large models. Our central hypothesis is that, by using MLDG to optimize only the low-rank LoRA adapters inside a frozen SSL backbone and the AASIST back-end, we can steer learning toward the fundamental, domain-invariant acoustic cues that distinguish synthetic speech. Full fine-tuning, in contrast, suffers from a built-in dilemma: although its larger capacity could in principle discover such cues, when trained with conventional objectives on a finite set of known attacks its expressivity is overwhelmingly spent memorizing superficial, domain-specific artefacts, leading to poor zero-shot transfer. 

Our approach imposes \textbf{both} a structural constraint (LoRA) and an objective-level constraint (MLDG), biasing optimization toward a more transferable solution space that full fine-tuning rarely reaches in practice—a claim borne out by the results in Sec.~\ref{sec:results}.

\begin{figure*}[!t]
    \centering
    \includegraphics[scale=0.21]{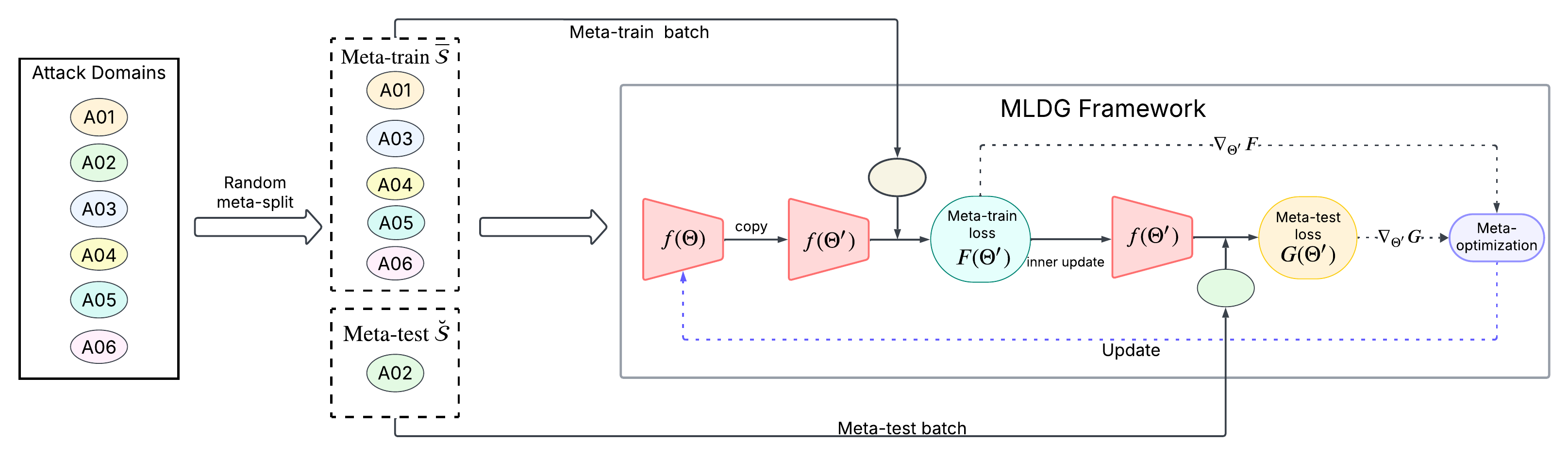} 
    \caption{Methodological framework for MLDG-based spoof detection. The depicted source domains (A01--A06) each contain their specific spoof type and corresponding bonafide audio. These are randomly split into meta-train and meta-test subsets to simulate domain shifts. While the diagram illustrates the core update logic (model cloning, inner update, loss computation) for one such meta-train/test evaluation, our full MLDG iteration (Algorithm~\ref{alg:fomldg}) aggregates gradients from multiple ($N_{pairs}$, e.g., 5) such internal pathways to update the main model ($\Theta$) for \textbf{generalization to unseen attacks}.}
    \label{fig:mldg_diagram}
\end{figure*}

Our study yields the following key findings and contributions:

\begin{itemize}
    \item Combining low‑rank LoRA adapters with MLDG reduces out‑of‑domain EER from 8.8\,\% to 5.3\,\% across five unseen‑attack corpora, while updating only 3.6M parameters ($\sim$1\,\% of the backbone) and halving seed‑to‑seed variance.
    
    \item With \emph{identical} adapter capacity, replacing ERM with MLDG lowers mean EER by 2--3 percentage points, showing that the meta‑learning objective—rather than added capacity—is the primary driver of generalization.
    
    \item Systematically varying LoRA rank from 2 to 16 reveals diminishing returns once MLDG is applied: algorithmic choice outweighs adapter size, suggesting that future work should prioritize advanced learning objectives over simply scaling parameter‑efficient modules.
\end{itemize}

\section{Background and related work}
\label{sec:background}

This section reviews prior work on domain generalization, speech deepfake detection, and approaches to parameter‑efficient model adaptation.

\textbf{Domain generalization}  
Generalizing machine‑learning models to unseen domains while preserving task performance is a long‑standing challenge \citep{Zhou_2022}.  
Existing strategies include data augmentation, domain alignment, and robust feature learning.  
Among them, \emph{meta‑learning} is a promising direction, with methods such as Meta‑Learning Domain Generalization (MLDG) \citep{Li2017LearningTG}, MetaReg \citep{Balaji_2018}, Model‑Agnostic Learning of Semantic Features (MASF) \citep{Dou_2019}, and Meta‑Variational Information Bottleneck (MetaVIB) \citep{Du_2020}.  
Specifically, MLDG partitions the source domains into \emph{meta‑train} and \emph{meta‑test} sets each iteration so that the resulting parameters perform well after adapting to any subset of domains.

Other approaches tackle evolving attacks and shifting data distributions.  
Continual‑learning techniques adapt models to new data while mitigating catastrophic forgetting \citep{Ma2021}.  
These methods often require access to the full dataset during updates.  
In contrast, meta‑learning aims for rapid adaptation from limited data.  
For example, Kukanov \emph{et al.}\ \cite{Kukanov_2024} study \emph{few‑shot} spoof detection with prototypical networks and MAML variants, and demonstrate that a few target samples can yield significant gains on unseen attacks.  
Our work differs from both continual learning and few‑shot adaptation by pursuing zero‑shot generalization with MLDG; we require no target‑domain samples for adaptation or testing.

\textbf{Speech spoof detection}  
Speech spoof detection—also called synthetic‑speech or deepfake‑audio detection—distinguishes bonafide speech from synthetically generated or manipulated (spoof) speech.  
Research has been accelerated by the ASVspoof challenge \emph{campaigns} \citep{Todisco2019ASVspoof2F,ASVspoof2021,Wang2024ASVspoof5C}, which provide standardized datasets and evaluation protocols.  
Early systems relied on engineered acoustic features, whereas later work adopted CNNs such as Light CNN (LCNN) \citep{lavrentyeva2019stcantispoofingsystemsasvspoof2019}.  
Architectures now incorporate large self‑supervised (SSL) models such as \texttt{Wav2Vec 2.0} \citep{Baevski_2020} or \texttt{HuBERT} \cite{HsuHubert} as powerful feature extractors.  
A prominent example is the Wav2Vec–AASIST model \citep{Tak2022Automatic,Jung_2022}, which pairs a \texttt{Wav2Vec 2.0} front-end with a graph‑attention‑based back-end.  
Although such systems achieve $<\!1\,\%$ EER on in‑domain test sets, generalizing to unseen spoofing attacks or different acoustic conditions remains challenging \citep{müller2024doesaudiodeepfakedetection}, motivating the need for domain‑generalization techniques.

\textbf{Parameter‑efficient fine‑tuning (PEFT)}  
Full fine‑tuning of a large pre‑trained model is both memory‑ and time‑intensive.  
PEFT methods update only a small subset of parameters.  
Among them, LoRA  \citep{hu2021loralowrankadaptationlarge} freezes the original weights and inserts trainable low‑rank matrices \(A\) and \(B\) with update \(\Delta W = AB\), reducing the trainable parameters by orders of magnitude—often to $<\!1\,\%$ of the original—thereby lowering memory use and accelerating training.  
LoRA has proved effective for large language models and, more recently, for speech spoof detectors: on \texttt{Wav2Vec 2.0} it achieves a \(198\times\) parameter reduction while matching full fine‑tuning performance \citep{wang2023lowrankadaptationmethodwav2vec2based}.  
Zhang \emph{et al.}\ \cite{zhang2023adaptivefakeaudiodetection} further show that LoRA improves robustness to unseen attacks.  
In this work we adopt LoRA as our PEFT mechanism and optimize its adapters with MLDG to obtain zero‑shot generalization.

\begin{table*}[t]
  \caption{A summary of the most relevant previous research used as inspiration for our work. TF denotes transformers.}
  \label{tab:previous_research}
  \centering
  \resizebox{\linewidth}{!}{%
  \begin{tabular}{l>{\raggedright\arraybackslash}p{4cm}lll}
    \toprule
    Paper & Description & Algorithms & Model & Uses eval data? \\
    \midrule
    \addlinespace[0.5em]
    Attack Agnostic \cite{Kawa2022} 
      & Pools data from different corpora; 5‑fold  CV 
      & ERM 
      & LCNN 
      & No \\
    \addlinespace[0.5em]
    
    Wav2Vec+AASIST \cite{Tak2022Automatic}
      & Coupling Wav2Vec 2.0 front-end with AASIST back-end 
      & ERM 
      & TF + GNN 
      & No \\
    \addlinespace[0.5em]

    Wav2Vec+LoRA \cite{zhang2023adaptivefakeaudiodetection}
      & Continual learning from three corpora with finetuned LoRA 
      & ERM 
      & TF + SENet, LoRA 
      & Yes \& No \\
    \addlinespace[0.5em]

    Adaptation with Meta-Learning \cite{Kukanov_2024}
      & Adapt Wav2Vec using ProtoMAML and ProtoNet 
      & Meta 
      & TF + GNN 
      & Yes \\
    \addlinespace[0.5em]

    \textbf{(Ours)} 
      & Finetunes LoRA parameters using MLDG; one corpus only 
      & Meta 
      & TF + GNN, LoRA 
      & No \\
    \addlinespace[0.5em]
    \bottomrule
  \end{tabular}%
  }
\end{table*}

\section{Proposed Method}
\label{sec:method}

Our approach combines meta-learning for domain generalization with parameter-efficient fine-tuning to improve the zero-shot generalization capability of speech spoof detectors against unseen attacks. Specifically, we use Meta-Learning Domain Generalization (MLDG) \citep{Li2017LearningTG} to optimize Low-Rank Adaptation (LoRA) \citep{hu2021loralowrankadaptationlarge} modules injected into a Wav2Vec 2.0-backbone \citep{Tak2022Automatic}. This strategy allows efficient adaptation focused on learning generalizable features by updating only a small, targeted set of parameters, while keeping the majority of the powerful pre-trained model's parameters frozen, thereby preserving pre-trained knowledge and potentially regularizing against overfitting to training domains \footnote{Code available at \url{https://github.com/Vanova/MLDGDeepfake}}. 

\subsection{Base Architecture}
\label{subsec:base_architecture}

We adopt a hybrid architecture that uses large self-supervised learning (SSL) models as frozen front ends, paired with a learnable back-end for binary classification. This design follows recent trends in deepfake detection, where SSL models pretrained on raw audio serve as powerful feature extractors \cite{Tak2022Automatic, Kukanov_2024}. The model comprises two main components:
\begin{itemize}
    \item \textbf{Front End:} We use either the Wav2Vec 2.0 \citep{Baevski_2020} or HuBERT \citep{HsuHubert} model as the front-end encoder. Both are transformer-based SSL models trained on large-scale raw speech corpora using masked prediction objectives. In our main experiments, we use the \texttt{Wav2Vec 2.0 XLSR-53} model, which is trained on 56,000 hours of multilingual audio across 53 languages and outputs 1024-dimensional embeddings. In an ablation study, we also evaluate the \texttt{HuBERT Base} model, trained on 960 hours of English LibriSpeech, which produces 768-dimensional representations. Regardless of the model, we extract the final-layer contextual embeddings and treat them as fixed features throughout training.   
    \item \textbf{Back End:} The back end is the AASIST spoofing countermeasure system \citep{Jung_2022}, which operates on the sequence of SSL embeddings. It uses spectro-temporal graph attention networks to model relationships across both time and feature dimensions, culminating in a binary classifier that outputs a spoof/bonafide prediction.
\end{itemize}
In our proposed method, the SSL front end remains frozen to preserve the generalizable speech representations learned during pretraining. We introduce lightweight LoRA adapters into selected attention and projection layers of the SSL encoder (described in Section~\ref{subsec:lora_integration}) and train only these adapters along with the AASIST back end parameters.

\begin{figure}[!ht]
 \centering
 \includegraphics[scale=0.21]{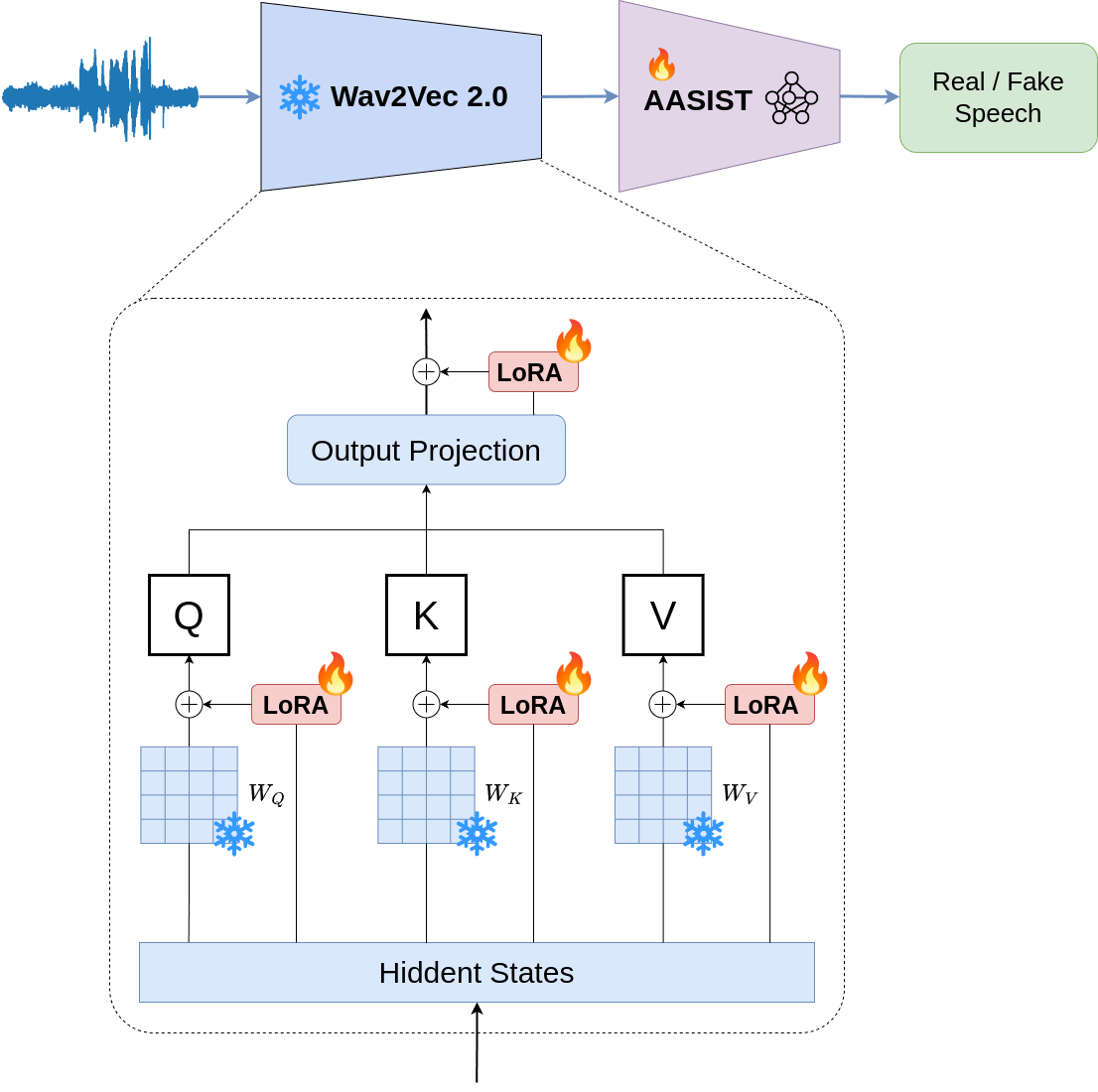}
 \caption{LoRA adapters (blue) inserted in the self-attention heads of Wav2Vec–AASIST. Only the adapters and the AASIST back-end are trainable.}
 \label{fig:lora_modules}
\end{figure}

\subsection{LoRA Integration}
\label{subsec:lora_integration}

To enable parameter-efficient adaptation of the Wav2Vec 2.0 front end, we employ Low-Rank Adaptation (LoRA) \citep{hu2021loralowrankadaptationlarge}. Trainable LoRA modules are injected into the query ($W_q$), key ($W_k$), value ($W_v$) and output-projection ($W_{\text{out}}$) linear layers of every self-attention block (Figure~\ref{fig:lora_modules}).

Each module approximates the update that full fine-tuning would apply to a pre-trained weight matrix $W$ with a product of two low-rank matrices, $\Delta W = AB$, where $A\!\in\!\mathbb{R}^{d\times r}$ and $B\!\in\!\mathbb{R}^{r\times m}$ and $r\!\ll\!d,m$. We sweep ranks $r\!\in\!\{2,4,8,16\}$; the LoRA scaling factor $\alpha_{\text{LoRA}}$ is fixed (\(\alpha_{\text{LoRA}}=2\)).

Because Wav2Vec 2.0 was pre-trained only on natural speech, its frozen weights~$W$ already capture the typical characteristics of bonafide audio. The LoRA adapters, guided by the MLDG objective (Section~\ref{subsec:mldg_optimization}), learn low-rank adjustments that fine-tune those features so that spoof artefacts become easier for the downstream AASIST classifier to separate; MLDG steers the adjustments toward cues that remain effective across unseen attack types. 

LoRA is particularly well-suited for this framework as it updates only about \(1\,\%\) of the backbone parameters even at rank 16, keeps the bulk of the pre-trained bonafide representation intact, and implicitly regularizes the model by restricting updates to a low-rank sub-space. During training we therefore update only the LoRA matrices \(A\) and \(B\) and the AASIST back-end; all other backbone weights remain fixed.

\subsection{MLDG Optimization}
\label{subsec:mldg_optimization}

We optimize the trainable parameters $\Theta$ (LoRA adapters + AASIST back-end) using the MLDG algorithm \citep{Li2017LearningTG}.

\textbf{Domain Definition:} We structure the training data into distinct domains based on the known attack types in the ASVspoof 2019 LA training set (A01--A06). Each domain $\mathcal{D}_k$ ($k = 1, \ldots, K$; $K = 6$) contains all spoofed utterances of the corresponding attack type $k$, along with a disjoint subset of the available bonafide utterances. The set of all source domains is denoted by $\mathcal{S} = \{\mathcal{D}_1, \ldots, \mathcal{D}_K\}$.

\textbf{Meta-Learning Procedure:} Each training iteration simulates a domain shift (Figure~\ref{fig:mldg_diagram}, Algorithm~\ref{alg:fomldg}):

\begin{enumerate}
    \item \textbf{Meta-Split:} Randomly partition domain indices $\{1, \ldots, K\}$ into meta-train $\overline{\mathcal{S}}$ and meta-test $\tilde{\mathcal{S}}$ sets (e.g., $|\tilde{\mathcal{S}}|=1$).

    \item \textbf{Meta-Train Update Simulation:} Create a clone of the model with parameters $\Theta' \leftarrow \texttt{clone}(\Theta)$. Compute the meta-train loss $F(\Theta')$ on mini-batches from $\overline{\mathcal{S}}$:
    \begin{equation} 
    \label{eq:mldg_f_theta}
    F(\Theta') = \frac{1}{|\overline{\mathcal{S}}|} \sum_{k \in \overline{\mathcal{S}}} \mathbb{E}_{(x,y) \sim \mathcal{D}_k} \left[ \ell(f_{\Theta'}(x), y) \right]
    \end{equation}
    Estimate the gradient $\nabla_{\Theta'} F(\Theta')$ and perform an inner update:
    \begin{equation}
    \label{eq:mldg_inner_update}
    \Theta' \leftarrow \Theta' - \alpha \nabla_{\Theta'} F(\Theta')
    \end{equation}

    \item \textbf{Meta-Test Loss Calculation:} Evaluate the updated $\Theta'$ on mini-batches from $\tilde{\mathcal{S}}$:
    \begin{equation}
    \label{eq:mldg_g_theta_prime}
    G(\Theta') = \frac{1}{|\tilde{\mathcal{S}}|} \sum_{k \in \tilde{\mathcal{S}}} \mathbb{E}_{(x,y) \sim \mathcal{D}_k} \left[ \ell(f_{\Theta'}(x), y) \right]
    \end{equation}
\end{enumerate}

\begin{algorithm}[tb]
   \caption{First-order MLDG algorithm (as implemented)}
   \label{alg:fomldg}
\begin{algorithmic}[1]
   \STATE \textbf{Input:} domains $\mathcal{S}$
   \STATE \textbf{Initialize:} Model parameters $\Theta$; hyperparameters $\alpha, \beta, \gamma$
   \FOR{\texttt{iteration} = 1 \textbf{to} \texttt{maxIters}}
    \STATE \textbf{Split:} $(\overline{\mathcal{S}}, \tilde{\mathcal{S}}) \leftarrow \texttt{split}(\mathcal{S})$
    \STATE \textbf{Clone:} $\Theta' \leftarrow \texttt{clone}(\Theta)$
    \STATE \textbf{Meta-train:} Compute $F(\overline{\mathcal{S}}; \Theta')$ and update $\Theta'$: $\Theta' \leftarrow \Theta' - \alpha \nabla_{\Theta'} F(\overline{\mathcal{S}}; \Theta')$
    \STATE \textbf{Meta-test:} Evaluate $G(\tilde{\mathcal{S}}; \Theta')$
    \STATE \textbf{Meta-optimization (First-Order):}
    \[
       \Theta \;\leftarrow\; \Theta  
         \;-\; \gamma\,\left(
           \nabla_{\Theta'} F(\overline{\mathcal{S}}; \Theta') 
           \;+\; \beta\, \nabla_{\Theta'} G(\tilde{\mathcal{S}}; \Theta') 
         \right)
    \]
   \ENDFOR
\end{algorithmic}
\end{algorithm}

\textbf{First-Order Parameter Update:} The parameters $\Theta$ are updated using gradients from both the meta-train and meta-test phases. A full second-order MLDG update, as originally formulated \citep{Li2017LearningTG}, requires differentiating the meta-test loss $G(\Theta')$ with respect to the initial parameters $\Theta$. This involves backpropagating through the inner gradient update step ($\Theta' = \Theta - \alpha \nabla_{\Theta} F(\Theta)$), necessitating the computation of second derivatives (e.g., Hessian-vector products).

To maintain computational tractability given the size of our model, we employ a first-order approximation of MLDG, which avoids computing second derivatives through the inner update. Instead of differentiating the meta-test loss $G(\Theta')$ with respect to the original parameters $\Theta$ (which would require backpropagating through the update $\Theta' = \Theta - \alpha \nabla_\Theta F(\Theta)$), we compute gradients only with respect to $\Theta'$, treating it as fixed during meta-test loss evaluation.

This variant is similar to the First-Order MAML (FOMAML) approximation introduced in \citep{finn17a}, which reduces memory usage and computation by omitting second-order terms. While our implementation still involves cloning the model and computing separate gradients for meta-train and meta-test tasks, it avoids the high memory cost of second-order backpropagation. Thus, it offers a practical trade-off between fidelity to the original algorithm and computational feasibility. Empirically, such approximations have been shown to retain competitive performance in meta-learning tasks \citep{finn17a, nichol2018}.

Our final update rule is:
\begin{equation}
\label{eq:mldg_outer_update_main}
\Theta \leftarrow \Theta - \gamma \left( \nabla_{\Theta'} F(\overline{\mathcal{S}}; \Theta') + \beta \nabla_{\Theta'} G(\tilde{\mathcal{S}}; \Theta') \right)
\end{equation}
Here, $\beta$ is a hyperparameter balancing the meta-train and meta-test gradients. We use negative log-likelihood (NLL) as the base loss $\ell(\cdot, \cdot)$ for binary classification. Our goal is to promote robust, domain-invariant representations across unseen attack domains.

\begin{table*}[htbp]
\centering
\caption{Summary of datasets (corpora) used in the experiments. The horizontal line divides datasets used in training and evaluation. ``\#Attacks'' denotes the number of distinct spoofing algorithms; ``N/R'' indicates unavailable counts. The ASVspoof 2021 DF subset combines audio from ASVspoof2019-LA and Voice Conversion Challenge 2018/2020 under various compression conditions.}
\label{tab:datasets}
\footnotesize
\setlength{\tabcolsep}{4pt}
\begin{tabular}{lcccc}
\toprule
Dataset & Usage & \#Bonafide & \#Spoofed & \#Attacks \\
\midrule
ASVSpoof 2019 LA Train \cite{Todisco2019ASVspoof2F} & Training & 2,580 & 22,800 & 6 \\
ASVSpoof 2019 LA Dev \cite{Todisco2019ASVspoof2F} & Validation & 2,548 & 22,296 & 6 \\ 
\cmidrule(r){1-5}
ASVSpoof 2019 LA Eval \cite{Todisco2019ASVspoof2F} & Evaluation & 7,355 & 63,882 & 13 \\
ASVSpoof 2021 LA \cite{ASVspoof2021} & Evaluation & 14,816 & 133,360 & 13 \\
ASVSpoof 2021 DF \cite{ASVspoof2021} & Evaluation & 14,869 & 519,059 & 100+ \\
InTheWild \cite{müller2024doesaudiodeepfakedetection} & Evaluation & 19,963 & 11,816 & N/R \\
FakeAVCeleb \cite{khalid2022fakeavcelebnovelaudiovideomultimodal} & Evaluation & 10,209 & 11,335 & 1 \\
ASVSpoof 5 LA Eval \cite{Wang2024ASVspoof5C} & Evaluation & 138,688 & 542,086 & 32 \\
\bottomrule
\end{tabular}
\end{table*}
\section{Experimental Setup}
\label{sec:exp_setup}
\subsection{Datasets}
Data used to train and evaluate the trained models are summarized in Table~\ref{tab:datasets}. To train and validate each model, we employed the Logical Access (LA) partition of the ASVspoof2019 dataset \cite{Todisco2019ASVspoof2F}, which is one of the most commonly used benchmarks in speech deepfake detection research. The dataset consists of English speech recordings derived from the VCTK corpus, covering both bonafide and spoofed utterances. The spoofed data is generated by a variety of text-to-speech (TTS) and voice conversion (VC) algorithms, collectively referred to as "attacks". The training and development subsets share six known attack types, labeled A01-A06. The evaluation set features additional attacks, labeled A07-A19, many of which differ from those seen in training.

To assess the generalization capability of our methods, we evaluate the models against both in-domain and out-of-domain datasets. The in-domain datasets include the evaluation sets of ASVSpoof 2019 LA, ASVSpoof 2021 LA and ASVSpoof 2021 DF \cite{ASVspoof2021}. The out-of-domain datasets are represented by In-The-Wild \cite{müller2024doesaudiodeepfakedetection} and FakeAVCeleb \cite{khalid2022fakeavcelebnovelaudiovideomultimodal}. Additionally, we evaluate our models against the recent ASVSpoof 5 \cite{Wang2024ASVspoof5C} evaluation set.
\subsection{Training Strategy}

\textbf{Baseline models.} We adopt Wav2Vec2-AASIST and Wav2Vec2-AASIST* as our primary baseline systems, building on methods previously described in \cite{Kukanov_2024}. The Wav2Vec2-AASIST model use Wav2Vec 2.0 (XLSR-53) as a front end and AASIST as the back end; both components are jointly trained. In the Wav2Vec-AASIST* variant, the Wav2Vec 2.0 component is frozen throughout training, and only the AASIST parameters are updated. Both versions are trained using standard empirical risk minimization (ERM) with a batch size of 16 utterances, in which we simply use the combined training data (bonafides + all spoofed attacks (A01-A06)) as a single dataset. For each of the baseline and all the consequent models we use the AdamW \cite{loshchilov2019decoupledweightdecayregularization} optimizer with a cyclic learning rate scheduler, oscillating between a minimum of $1 \times 10^{-7}$ and a maximum of $1 \times 10^{-5}$ over each cycle. We optimize a negative log-likelihood loss over two-class (bonafide vs spoof) log-softmax outputs. As a stopping criterion, we monitor validation performance using the ASVSpoof 2019 LA validation subset, and terminate training if no improvement occurs for more than 10 epochs, saving the best-performing checkpoint for final testing.

\textbf{ERM LoRA models.} In the ERM LoRA versions, we inject LoRA adapters into the self-attention layers of the SSL backbone (either Wav2Vec or HuBERT), covering the query, key, value, and output projection modules. The original backbone are frozen, and only the LoRA modules and the AASIST back end are trainable. We experiment with LoRA ranks $r \in {2,4,8,16}$ for our main experiments with Wav2Vec 2.0, while fixing the scaling factor to $\alpha=2$. All models use the same optimizer, batch size, scheduler, and loss function as the baseline models.

\begin{table*}[!htbp]
  \centering
  \caption{Comparison of baselines and best LoRA models (rank 16) trained with
           ERM and MLDG. Results are EER (\%) as mean ± std over 5 seeds.
           Bold indicates the best performance in each column.}
  \label{tab:main_results}

  \footnotesize
  \setlength{\tabcolsep}{2.0pt}
  \renewcommand{\arraystretch}{1.25}

  \begin{adjustbox}{max width=\linewidth, center}
  \begin{tabular}{@{}l c *{7}{r@{$\,\pm\,$}l}@{}}
    \toprule
    \multicolumn{1}{c}{\multirow{2}{*}{Model}} &
    \multicolumn{1}{c}{\multirow{2}{*}{\shortstack[c]{Trainable\\Params}}} & 
    \multicolumn{2}{c}{\multirow{2}{*}{ASV19:LA}} & 
    \multicolumn{2}{c}{\multirow{2}{*}{ASV21:LA}} &
    \multicolumn{2}{c}{\multirow{2}{*}{ASV21:DF}} &
    \multicolumn{2}{c}{\multirow{2}{*}{InTheWild}} &
    \multicolumn{2}{c}{\multirow{2}{*}{FakeAVCeleb}} &
    \multicolumn{2}{c}{\multirow{2}{*}{ASV5}} &
    \multicolumn{2}{c}{\multirow{2}{*}{Avg.}} \\
      & & \multicolumn{2}{c}{} & \multicolumn{2}{c}{} &
          \multicolumn{2}{c}{} & \multicolumn{2}{c}{} &
          \multicolumn{2}{c}{} & \multicolumn{2}{c}{} &
          \multicolumn{2}{c}{} \\
    \midrule
    Wav2Vec\textminus AASIST ERM & 317.8M
      & \textbf{0.19} & \textbf{0.07}
      & 5.88 & 2.26
      & 5.13 & 1.02
      & 12.15 & 2.79
      & 6.27 & 1.47
      & 23.38 & 2.52
      & 8.84 & 0.71 \\
    \addlinespace

    % renamed row (ERM)
    Wav2Vec\textminus AASIST* ERM & 447K
      & 1.00 & 0.44
      & 8.04 & 1.45
      & 8.94 & 2.17
      & 25.42 & 1.64
      & 7.72 & 2.40
      & 22.19 & 7.40
      & 12.22 & 1.83 \\

    % new MLDG row – replace the placeholders
    Wav2Vec\textminus AASIST* MLDG & 447K
      &2.91 &1.53 
      &10.59  &3.62 
      &8.18  &0.83
      &19.08 &3.82
      &8.26 &5.15
      &17.95 &2.47
      &10.99 &2.50 \\
    \midrule

    LoRA (ERM, Rank 16) & 3.59M
      & 0.47 & 0.15
      & 7.66 & 3.83
      & 4.35 & 1.37
      & 11.05 & 6.27
      & 3.89 & 2.95
      & 20.05 & 4.36
      & 7.91 & 2.82 \\
    \addlinespace

    LoRA (MLDG, Rank 16) & 3.59M
      & 0.54 & 0.33
      & \textbf{4.86} & \textbf{1.02}
      & \textbf{3.99} & \textbf{0.46}
      & \textbf{6.81} & \textbf{0.81}
      & \textbf{1.48} & \textbf{1.08}
      & \textbf{14.10} & \textbf{0.39}
      & \textbf{5.30} & \textbf{0.37} \\
    \bottomrule
  \end{tabular}
  \end{adjustbox}
  \normalsize
\end{table*}
\textbf{MLDG LoRA models.} To improve generalization to unseen spoof types, we also explore LoRA-based models under a meta-learning regime rather than straightforward ERM. We adopt a first-order variant of Meta-Learning Domain Generalization (MLDG) that partitions the ASVspoof 2019 LA training set into six domain-specific subsets $\{\mathcal{D}_1,\dots,\mathcal{D}_6\}$ based on spoof attack type. At each iteration, we sample a mini-batch of 3 utterances from each domain, forming a six-domain meta-batch of 18 utterances. These domains are then split randomly at each iteration into meta-train and meta-test domains. A clone of the model parameters $\Theta$ is updated using the meta-train batch and the AdamW optimizer (inner learning rate $\alpha=0.001$), resulting in adapted parameters $\Theta'$. The meta-test loss is computed on $\Theta'$, and its gradient is scaled by a hyperparameter $\beta = 0.5$ before being accumulated into the main model's gradient buffer. The outer update is applied to $\Theta$ using AdamW. As with the ERM LoRA setup, only the LoRA and AASIST parameters are trainable, while the SSL backbone remains frozen.   

\textbf{Front-end Backbone Features.} To show that MLDG is backbone agnostic, we additionally perform an ablation experiment using the same architecture but with a HuBERT-Base backbone instead of Wav2Vec 2.0. We refer to these models as HuBERT-AASIST and HuBERT-AASIST*, as HuBERT-AASIST* mirrors Wav2Vec-AASIST* in keeping the front-end frozen and updating only the AASIST back end.  A complete list of hyperparameters used in our experiments is provided in the appendices.

\section{Results}
\label{sec:results}
\textbf{Main comparison.}  
Table \ref{tab:main_results} considers five variants.  
The fully fine-tuned baseline, which updates all 318 M parameters of Wav2Vec-AASIST, averages \(8.84\%\pm0.71\) EER across the six evaluation corpora.  
If the SSL front end is frozen and only the 447K AASIST weights are trained with ERM, performance falls to \(12.22\%\pm1.83\); training that same AASIST-only model with first-order MLDG brings a mild improvement to \(10.99\%\pm2.50\), showing that meta-learning helps even when very few parameters are free.  

Replacing the full fine-tune updates with 3.6 M rank-16 LoRA adapters already changes the picture: under ERM those adapters reach \(7.91\%\pm2.82\) updating only about 1\% of the parameters compared to full fine-tuning.  
Applying MLDG to the \emph{same} adapters pushes the mean EER down further to \textbf{5.30 \%} and, at the same time, reduces seed-to-seed variance from 2.82 to \textbf{0.37} percentage points.  
In short, LoRA delivers parameter efficiency and a baseline gain over full fine-tuning, whereas the MLDG objective supplies the decisive jump in generalization and stability.

\begin{table*}[!htbp]
  \centering
  \caption{Performance sweep for different LoRA configurations. Each cell shows EER (\%) mean\,$\pm$\,std
           over 5 seeds. For each LoRA rank and dataset, the lower EER between ERM and MLDG is bold-faced.}
  \label{tab:lora_sweep}

  \footnotesize
  \setlength{\tabcolsep}{2.5pt}
  \renewcommand{\arraystretch}{1.25}

  \begin{adjustbox}{max width=\linewidth, center} 
 
  \begin{tabular}{@{}l c c *{7}{r@{$\,\pm\,$}l}@{}}
    \toprule
  
    \multicolumn{1}{c}{\multirow{2}{*}{Method}} &
    \multicolumn{1}{c}{\multirow{2}{*}{\shortstack[c]{Trainable\\Params}}} &
    \multicolumn{1}{c}{\multirow{2}{*}{Rank}} &
    \multicolumn{2}{c}{\multirow{2}{*}{ASV19:LA}} &
    \multicolumn{2}{c}{\multirow{2}{*}{ASV21:LA}} &
    \multicolumn{2}{c}{\multirow{2}{*}{ASV21:DF}} &
    \multicolumn{2}{c}{\multirow{2}{*}{InTheWild}} &
    \multicolumn{2}{c}{\multirow{2}{*}{FakeAVCeleb}} &
    \multicolumn{2}{c}{\multirow{2}{*}{ASV5}} &
    \multicolumn{2}{c}{\multirow{2}{*}{Avg.}} \\

    & & & \multicolumn{2}{c}{} & \multicolumn{2}{c}{} & \multicolumn{2}{c}{} & \multicolumn{2}{c}{} & \multicolumn{2}{c}{} & \multicolumn{2}{c}{} & \multicolumn{2}{c}{} \\
    \midrule
    ERM  & \multirow{2}{*}{0.84M} & \multirow{2}{*}{2} & 
      0.69 & 0.11 & 7.67 & 3.66 & 5.10 & 1.16 & 17.02 & 3.89 & 8.94 & 3.66 & 20.28 & 3.96 & 9.95 & 1.51 \\
    MLDG &       &       & 
      \textbf{0.40} & \textbf{0.14} & \textbf{4.66} & \textbf{0.40} & \textbf{4.42} & \textbf{0.49} & \textbf{10.39} & \textbf{1.50} & \textbf{3.80} & \textbf{1.49} & \textbf{15.20} & \textbf{0.79} & \textbf{6.29} & \textbf{0.33} \\
    \addlinespace

    ERM  & \multirow{2}{*}{1.23M} & \multirow{2}{*}{4} &
      0.53 & 0.15 & 7.22 & 0.98 & 4.30 & 0.38 & 13.10 & 2.62 & 7.81 & 4.48 & 18.14 & 3.77 & 8.52 & 1.40 \\
    MLDG &       &       &
      \textbf{0.49} & \textbf{0.07} & \textbf{4.77} & \textbf{0.89} & \textbf{4.14} & \textbf{0.59} & \textbf{7.28} & \textbf{0.46} & \textbf{2.03} & \textbf{0.91} & \textbf{14.48} & \textbf{0.63} & \textbf{5.53} & \textbf{0.14} \\
    \addlinespace

    ERM  & \multirow{2}{*}{2.02M} & \multirow{2}{*}{8} &
      0.65 & 0.51 & \textbf{5.34} & \textbf{1.48} & \textbf{3.79} & \textbf{0.71} & 12.99 & 3.77 & 2.85 & 1.52 & 16.17 & 2.60 & 6.97 & 0.63 \\
    MLDG &       &       &
      \textbf{0.48} & \textbf{0.14} & 5.62 & 0.65 & 4.57 & 0.97 & \textbf{7.66} & \textbf{1.34} & \textbf{1.57} & \textbf{1.32} & \textbf{13.51} & \textbf{0.72} & \textbf{5.57} & \textbf{0.26} \\
    \addlinespace

    ERM  & \multirow{2}{*}{3.59M} & \multirow{2}{*}{16} &
      \textbf{0.47} & \textbf{0.15} & 7.66 & 3.83 & 4.35 & 1.37 & 11.05 & 6.27 & 3.89 & 2.95 & 20.05 & 4.36 & 7.91 & 2.82 \\
    MLDG &       &       &
      0.54 & 0.33 & \textbf{4.86} & \textbf{1.02} & \textbf{3.99} & \textbf{0.46} & \textbf{6.81} & \textbf{0.81} & \textbf{1.48} & \textbf{1.08} & \textbf{14.10} & \textbf{0.39} & \textbf{5.30} & \textbf{0.37} \\
    \bottomrule
  \end{tabular}%
  \end{adjustbox}
  \normalsize
\end{table*}

\begin{table*}[t]
  \centering
  \caption{Comparison of HuBERT baselines and best LoRA models (rank 16) trained with
           ERM and MLDG. Results are EER (\%) as mean\,$\pm$\,std over 5 seeds.
           Bold indicates the best performance in each column.}
  \label{tab:hubert_results}

  \footnotesize 
  \setlength{\tabcolsep}{2.0pt}
  \renewcommand{\arraystretch}{1.25}

  \begin{adjustbox}{max width=\linewidth, center}
  \begin{tabular}{@{}l c *{7}{r@{$\,\pm\,$}l}@{}}
    \toprule

    \multicolumn{1}{c}{\multirow{2}{*}{Model}} &
    \multicolumn{1}{c}{\multirow{2}{*}{\shortstack[c]{Trainable\\Params}}} & 
    \multicolumn{2}{c}{\multirow{2}{*}{ASV19:LA}} & 
    \multicolumn{2}{c}{\multirow{2}{*}{ASV21:LA}} &
    \multicolumn{2}{c}{\multirow{2}{*}{ASV21:DF}} &
    \multicolumn{2}{c}{\multirow{2}{*}{InTheWild}} &
    \multicolumn{2}{c}{\multirow{2}{*}{FakeAVCeleb}} &
    \multicolumn{2}{c}{\multirow{2}{*}{ASV5}} &
    \multicolumn{2}{c}{\multirow{2}{*}{Avg.}} \\

      & & \multicolumn{2}{c}{} & \multicolumn{2}{c}{} & \multicolumn{2}{c}{} & \multicolumn{2}{c}{} & \multicolumn{2}{c}{} & \multicolumn{2}{c}{} & \multicolumn{2}{c}{} \\
    \midrule
    HuBERT\textminus AASIST & 96.3M			
    & \textbf{0.85} & \textbf{0.20}
    & \textbf{4.87} & \textbf{0.82}
    & \textbf{9.58} & \textbf{0.76}
    & 15.64 & 2.06
    & 13.25 & 6.52
    & 26.00 & 3.41
    & 11.70 & 1.17 \\
    \addlinespace

    HuBERT\textminus AASIST* & 447K	
      & 5.13 & 1.03
      & 10.17 & 2.77
      & 14.19 & 0.68
      & 23.65 & 1.59
      & 22.94 & 3.90
      & 25.23 & 0.38
      & 16.88 & 1.52 \\
    \midrule

    LoRA (ERM, Rank 16) & 1.59M
      & 0.97 & 0.11
      & 5.94 & 1.46
      & 10.70 & 0.73
      & 19.67 & 1.84
      & 13.08 & 6.66
      & 26.74 & 2.11
      & 12.85 & 1.26 \\
    \addlinespace

    LoRA (MLDG, Rank 16) & 1.59M
      & 1.07 & 0.26
      & 5.76 & 1.48
      & 11.47 & 2.11
      & \textbf{13.58} & \textbf{1.38}
      & \textbf{12.72} & \textbf{2.74}
      & \textbf{22.07} & \textbf{0.91}
      & \textbf{11.11} & \textbf{0.51} \\
    \bottomrule
  \end{tabular}%
  \end{adjustbox}
  \normalsize
\end{table*}
\textbf{Effect of LoRA rank.} Table~\ref{tab:lora_sweep} sweeps ranks 2, 4, 8, and 16 under both training objectives. 
For every rank, the MLDG variant outperforms its ERM counterpart on all distribution-shifted corpora, showing that the inner–outer split—not adapter capacity—drives the generalization gain. 
Within the MLDG column, raising the rank from 2 (0.84 M parameters) to 4 nearly halves the average EER, from 6.29\,\% to 5.53\,\%, already surpassing the full fine-tuning baseline. 
Beyond that point the curve flattens: ranks 4, 8, and 16 obtain 5.53\,\%, 5.57\,\%, and 5.30\,\%, respectively, a spread of only 0.27 percentage points. 
The largest adapters (rank 16, 3.6 M parameters, $\approx$1 \% of the backbone) deliver the overall minimum, but the margin over rank 4 is small, suggesting that once meta-learning is applied, a moderate adapter is sufficient and additional rank yields diminishing returns.

\textbf{Stability across seeds.}
The variance columns in Tables~\ref{tab:main_results} and~\ref{tab:lora_sweep} reveal that first-order MLDG is also substantially more stable.  
For rank-16 adapters the ERM run spans nearly three percentage points across five random seeds, whereas MLDG confines that range to well under one point.  
We observe a similar variance reduction at every rank.

\begin{figure}[t]
    \centering
    \includegraphics[scale=0.15]{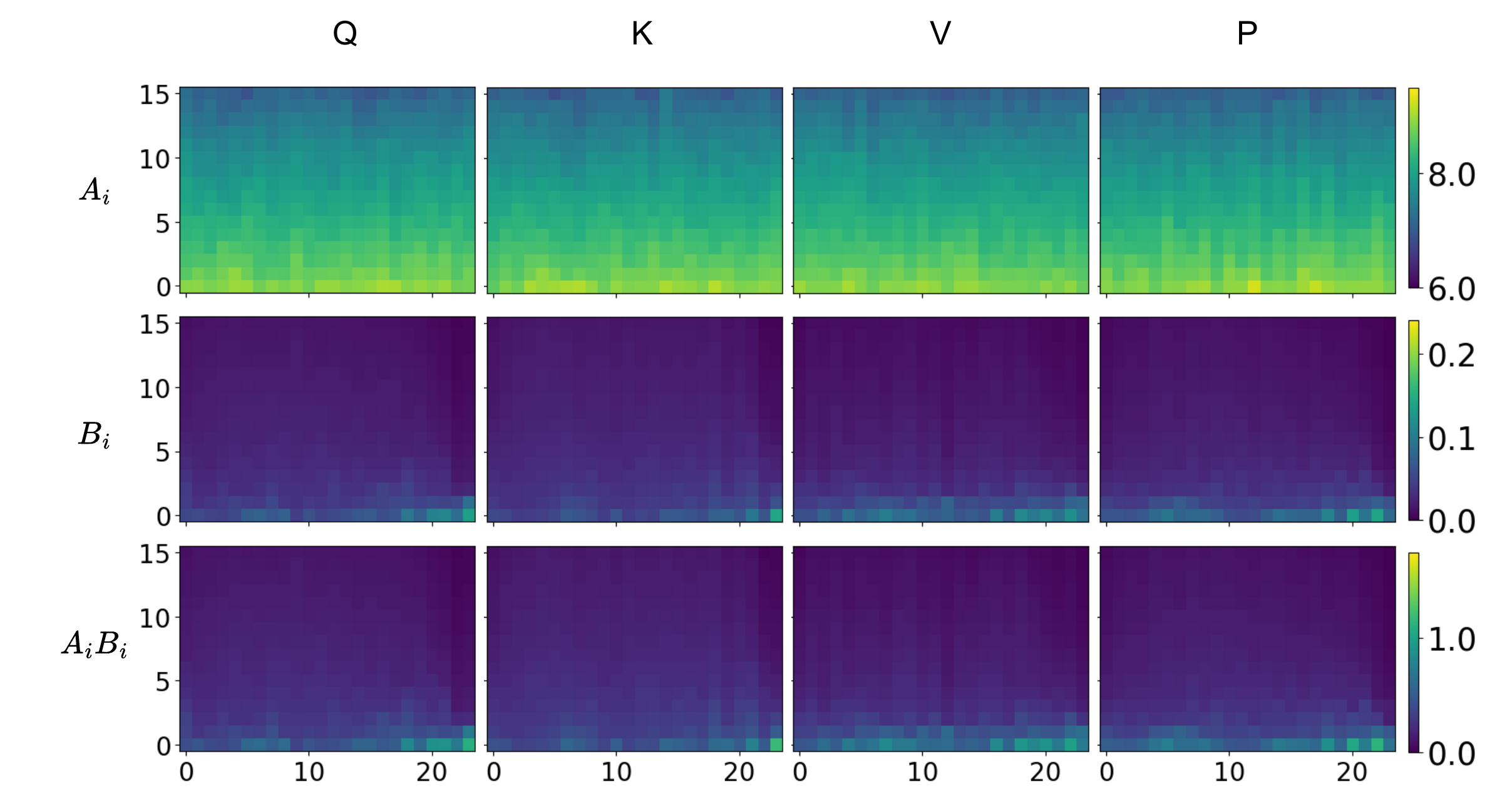}
    \vspace{1em}
    \includegraphics[scale=0.15]{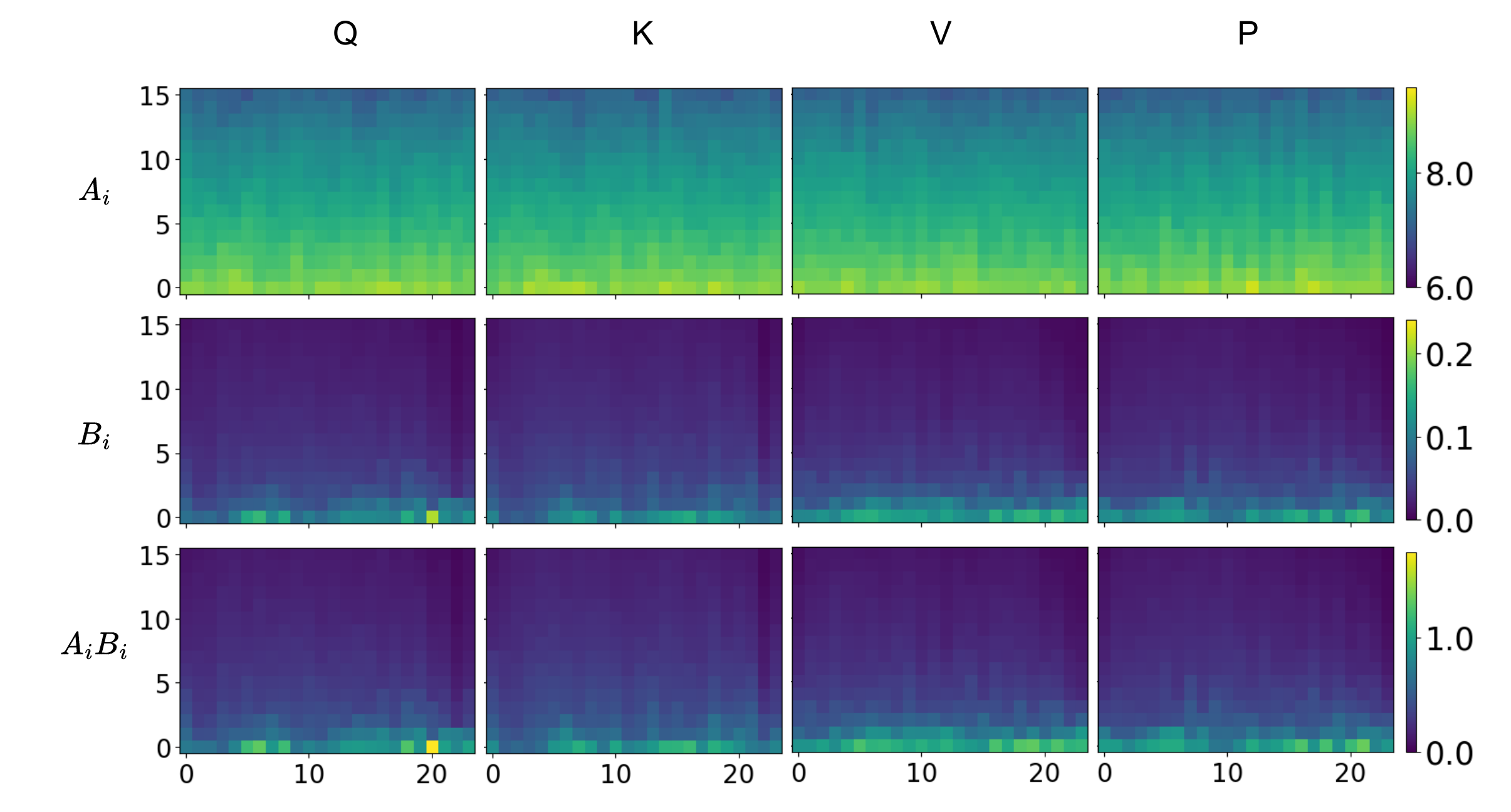}
    \caption{Singular values of LoRA weights $A_i$, $B_i$, and product $A_iB_i$ for rank 16, across 24 layers of Wav2Vec backbone. These are trained with ERM (top) and MLDG (bottom) methods from Tab.~\ref{tab:main_results}. The singular values are indexed on the y-axis and backbone layers on the x-axis for each heatmap.}
    \label{fig:erm_mldg_lora_svd}
\end{figure}
\textbf{Dataset-specific trends.}
A closer look shows that full fine-tuning remains unmatched on ASVspoof 2019 LA (0.19 \%), the evaluation set most similar to the training distribution, but suffers dramatic degradation on In-The-Wild and ASVspoof 5.  
Conversely, MLDG–LoRA sacrifices a fraction of a percentage point on the in-distribution set yet cuts error rates by factors of two to three on the corpora with the largest distribution shift, illustrating the usual trade-off between in-domain optimization and out-of-domain robustness.

\textbf{Alternative front end (HuBERT).} Table~\ref{tab:hubert_results} repeats the study with a 96M-parameter HuBERT front end.  
Fully fine-tuning serves as a 96M-parameter reference and yields an average EER of $11.70\%\pm1.17$.  
Freezing HuBERT’s weights degrades performance to $16.9\%$, but adding just 1.6 M rank-16 LoRA parameters under ERM already recovers much of the loss, reaching $12.85\%\pm1.26$—only one percentage point above full fine-tuning while training \(\approx\!60\times\) fewer weights.  
Applying first-order MLDG to the \emph{same} adapters delivers a further drop to \textbf{11.11\,\%} and cuts the seed-to-seed standard deviation to \textbf{0.51} pp.  
Thus, even on a weaker SSL backbone, LoRA gives a highly parameter-efficient alternative to full fine-tuning, and MLDG supplies an additional, variance-reducing boost that closes most of the remaining gap.

\textbf{Singular values of LoRA weights.} In Fig.~\ref{fig:erm_mldg_lora_svd}, the singular values of LoRA weights are analyzed, trained with the ERM or MLDG approaches in Table~\ref{tab:main_results}. The singular values of $A_i$ spread over multiple dimensions (highest intensity heatmap). This suggests that $A_i$ captures significant variance and retains more of the input rank/information. The middle plot $B_i$ is significantly darker, with most values close to zero. This implies $B_i$ has a much lower effective rank, acting as a bottleneck and detecting only specific features. The rank of $B_i$ is low in most positions of Wav2Vec, and concentrated toward the first few components. The final product $A_iB_i$ shows a pattern similar to $B_i$, though slightly richer. The structure of $A_iB_i$ suggests most of the learning signal comes from a small number of singular directions — beneficial for generalization and efficiency. Layer-wise characteristics show that some layers contribute more than others to the target task adaptation. Using MLDG, LoRA weights are more adapted and activate more singular values. Therefore, it needs more expressive power, possibly due to the domain shift training with meta-learning.

% \begin{figure*}[!ht]
%     \centering
%     \begin{minipage}{0.48\textwidth}
%         \centering
%         \includegraphics[scale=0.15]{erm_lora_weights.pdf}
%         % \caption{LoRA singular values for matrices $A_i$, $B_i$, and $A_iB_i$.}
%     \end{minipage}
%     \hfill
%     \begin{minipage}{0.48\textwidth}
%         \centering
%         \includegraphics[scale=0.15]{mldg_lora_weights.pdf}
%         % \caption{LoRA singular values for matrices $A_i$, $B_i$, and $A_iB_i$.}
%     \end{minipage}
%     \caption{Singular values of LoRA weights $A_i$, $B_i$, and product $A_iB_i$ for rank 16, across 24 layers of Wav2Vec backbone. These are trained with ERM (left) and MLDG (right) methods from Tab. \ref{tab:main_results}. The singular values are indexed in the y-axis and backbone layers in the x-axis for each heatmap.}
%     \label{fig:erm_mldg_lora_svd}
% \end{figure*}

\section{Conclusion}
\label{sec:conclusion}

This work investigated zero-shot adaptation for speech-spoof detection.  
By inserting Low-Rank Adaptation (LoRA) modules into a Wav2Vec 2.0–AASIST backbone and optimizing them with first-order MLDG, we lowered the average EER on six evaluation corpora with unseen attacks from 8.84 \% (full fine-tune) to 5.30 \%, updating only $\approx 1$ \% of the backbone parameters.  
The same rank-16 adapters trained with ERM achieved 7.91 \%, so the meta-learning signal, rather than parameter count, drives the improvement; it also decreases seed-to-seed variance from 2.82 pp to 0.37 pp.  
Hence, LoRA brings parameter efficiency, and MLDG supplies the generalization boost. By improving generalizable speech deepfake detection, this research helps strengthen defenses against audio misinformation and supports the integrity of digital communication.

\textbf{Limitations \& future work.}  
Our evaluation is English-only, trained on ASVspoof 2019-LA, and restricted to Transformer backbones plus first-order MLDG; we did not study cross-lingual transfer, larger or diffusion-based training sets, fairness across speaker traits, or adversarially crafted spoofs.  Ongoing work broadens the data (multilingual, DiffSSD), benchmarks additional DG objectives, explores non-Transformer encoders, and analyses the spoof artefacts captured by learned LoRA directions.

\section*{Acknowledgments}
VH was partially supported by the Jane and Aatos Erkko
Foundation. The authors acknowledge CSC, IT Center for
Science, Finland, for computational resources.

\bibliography{arxiv_ver}

\appendix
\section{Experimental Details}

\begin{table}[H]
\centering
\scriptsize  % Smaller font
\setlength{\tabcolsep}{4pt}  % Tighter column spacing
\renewcommand{\arraystretch}{0.95}  % Reduce vertical padding
\caption{Key hyperparameters.  LR = learning rate.}
\label{tab:hyperparameters}
\begin{tabularx}{\linewidth}{@{}l X@{}}
\toprule
\textbf{Hyperparameter} & \textbf{Value / Description} \\
\midrule
\multicolumn{2}{@{}l}{\textit{Outer-loop training (ERM and MLDG)}} \\
Optimizer & AdamW \\
Base LR & $1\!\times\!10^{-5}$ \\
LR schedule & Cyclic, triangular; min $1\!\times\!10^{-7}$, max $1\!\times\!10^{-5}$, step size $=12$ epochs \\
Early-stopping metric & \texttt{val\_eer} (ASV19 LA dev) \\
Patience & 10 epochs \\
Random seeds & $\{999,2023,555,123,42\}$ \\
ERM batch size & 16 utterances \\
\midrule
\multicolumn{2}{@{}l}{\textit{MLDG-specific}} \\
Inner-loop optimizer & AdamW \\
Inner-loop LR & $\alpha=0.001$ \\
Meta-test scaling & $\beta=0.5$ \\
Meta-batch composition & $K{=}6$ domains $\times\,3$ utt./domain $=18$ utt. \\
Meta-test domains per split & $|\tilde{\mathcal{S}}|=1$ \\
Inner gradient steps & 1 \\
\midrule
\multicolumn{2}{@{}l}{\textit{LoRA-specific}} \\
Rank $r$ & $\{2,4,8,16\}$ \\
Scaling $\alpha_{\text{LoRA}}$ & 2 \\
Target modules & $W_q$, $W_k$, $W_v$, $W_{\text{out}}$ in every self-attention block \\
\bottomrule
\end{tabularx}
\end{table}

Table~\ref{tab:hyperparameters} lists the hyper-parameters shared by all
experiments.  Unless stated otherwise, every model—baseline, ERM LoRA,
and MLDG LoRA—runs with the same optimiser, schedule, and early-stopping
rule.  For MLDG we use a single inner step of size
$\alpha=0.001$ and scale the meta-test gradient by $\beta=0.5$.
LoRA adapters are inserted into the query, key, value, and
output-projection projections of every self-attention block; rank
$r\in\{2,4,8,16\}$, with $\alpha_{\text{LoRA}}{=}2$ held fixed.

\begin{table}[H]
\scriptsize
\centering
\caption{Comparison of computational costs and resource usage during training on a single NVIDIA RTX 6000 Ada GPU (unless specified otherwise). Lower is better for GPU memory and epoch time; higher is better for throughput.}
\label{tab:computational_costs}
\begin{tabularx}{\linewidth}{@{} >{\raggedright\arraybackslash}X rrrr @{}} 
\toprule
\textbf{Method} & \textbf{Params} & \textbf{GPU Mem.} & \textbf{Time/Epoch} & \textbf{Thr'put}  \\
                & \textbf{(M)}    & \textbf{(GB) $\downarrow$} & \textbf{(min) $\downarrow$} & \textbf{(spl/s) $\uparrow$} \\ 
\midrule
Wav2Vec-AASIST          & 317.8 & 19.75 & 16.70 & 0.52 \\
Wav2Vec-AASIST* ERM     &   0.45&  8.82 &  8.09 & 0.99 \\
\midrule
ERM LoRA ($r{=}2$)      & 0.84  & 11.86 & 11.41 & 0.72 \\
ERM LoRA ($r{=}4$)      & 1.23  & 11.88 & 11.41 & 0.72 \\
ERM LoRA ($r{=}8$)      & 2.02  & 11.89 & 11.40 & 0.72 \\
ERM LoRA ($r{=}16$)     & 3.59  & 11.93 & 11.42 & 0.72 \\
\midrule
MLDG LoRA ($r{=}2$)     & 0.84  & 17.59 & 43.21 & 0.21 \\
MLDG LoRA ($r{=}4$)     & 1.23  & 17.60 & 43.40 & 0.21 \\
MLDG LoRA ($r{=}8$)     & 2.02  & 17.48 & 43.86 & 0.21 \\
MLDG LoRA ($r{=}16$)    & 3.59  & 17.55 & 44.27 & 0.21 \\
\bottomrule
\end{tabularx}
\end{table}

\section{Computational costs}
Table~\ref{tab:computational_costs} summarizes memory, runtime, and
throughput for every training configuration.
Freezing the $318\,\text{M}$-parameter backbone and fine-tuning only the
$3$–$4\,\text{M}$ LoRA adapters ($r{=}16$) lowers peak GPU memory from
19.8 GB to 11.9 GB (\(\approx 40\%\!\downarrow\)) and roughly halves the
epoch time, while slightly \emph{increasing} outer-step throughput.
Replacing ERM with first-order MLDG on the same adapters multiplies epoch
time by \(\approx 4\times\) and raises memory to 17.6 GB, because each
outer step evaluates five meta-train/meta-test pairs—cloning the model
and running ten forward/backward passes—before a single update on 18 new
utterances.  The added compute is worthwhile: MLDG-trained adapters
deliver the strongest cross-domain accuracy and the lowest seed-to-seed
variance shown in the main results table while leaving all 318 M backbone
weights untouched.

% ----------------------------------------------------------

\section{Additional experimental results}

\begin{figure}[H]
  \centering
  \includegraphics[width=.8\linewidth]{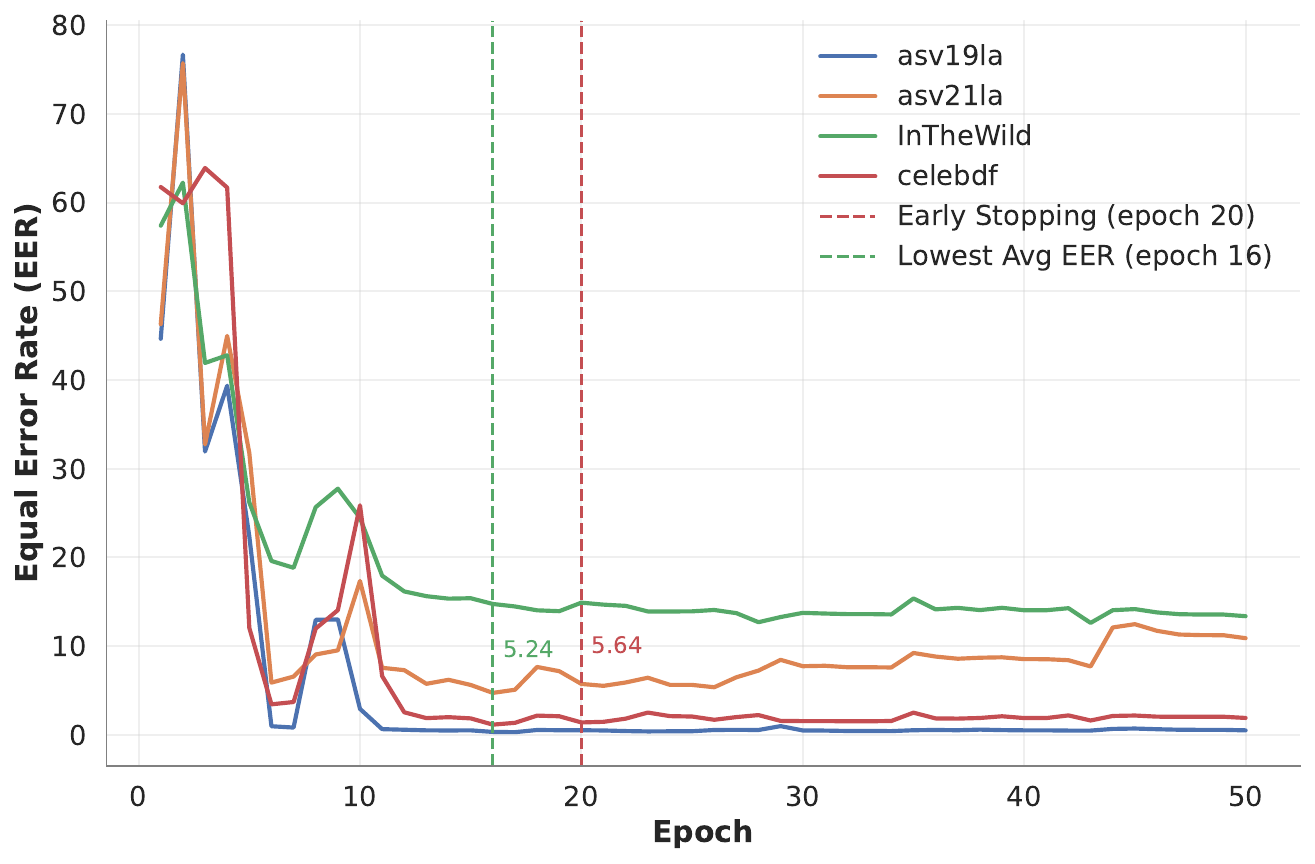}
  \caption{EER versus epoch for four evaluation sets.
           Red dashed line: epoch chosen by early stopping;
           green dashed line: epoch with the lowest average EER.}
  \label{fig:epochs}
\end{figure}

\paragraph{Early-stopping analysis.}
We trained one ERM LoRA model ($r{=}8$) for 50 epochs, saved a checkpoint
after each epoch, and evaluated it on four test sets
(ASV19 LA eval, ASV21 LA, In-The-Wild, FakeAVCeleb).
Figure~\ref{fig:epochs} shows the EER curves.
Early stopping terminated at epoch 20 (average EER 5.64\%), whereas the
global minimum, 5.24\%, occurred at epoch 16—0.40 pp lower.

\begin{figure}[H]
  \centering
  \includegraphics[width=.8\linewidth]{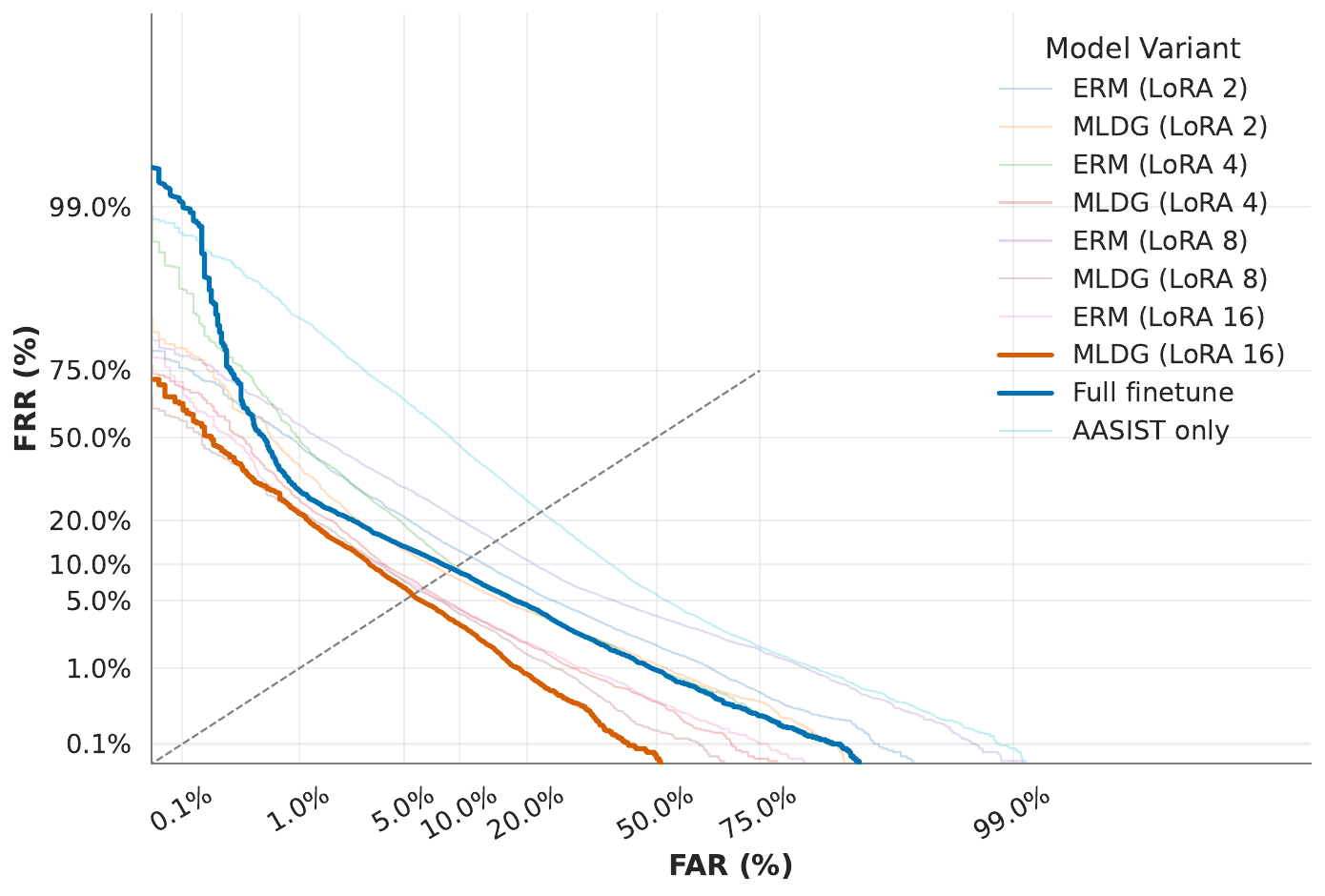}
  \caption{DET curves on \textit{In-The-Wild}.
           Horizontal axis = false-accept rate (FAR); vertical axis = false-reject rate (FRR).
           Highlighted curves: full fine-tune (blue) and LoRA $r{=}16$ MLDG (orange).}
  \label{fig:det_plot}
\end{figure}

\paragraph{Detection-error trade-off.}
Figure~\ref{fig:det_plot} plots DET curves on the
\textit{In-The-Wild} corpus.  For each variant we show the seed with the
lowest EER on that set.  The rank-16 LoRA model trained with MLDG
(outlined in orange) lies well below both its ERM counterpart and the
fully fine-tuned baseline across the operating range.  The nearly linear shape of the LoRA-16 + MLDG curve in probit space implies that the bonafide and spoof score distributions it produces are approximately Gaussian with similar variance — an indicator of good
calibration \cite{martin97b_eurospeech}.  In contrast, the full fine-tuned baseline exhibits a noticeable concavity, a signature of unequal class variances or heavy-tailed
score distributions, and thus weaker calibration. Smaller adapters
(ranks 2–8) sit between those extremes, mirroring the average-EER trends
in the table detailing the performance sweep for different LoRA configurations.  The AASIST-only ERM baseline forms the upper envelope, highlighting the value of even low-rank access to the frozen SSL features.

\end{document}